# Cascaded optical transparency in multimode-cavity optomechanical systems

Linran Fan[1], King Y. Fong[1], Menno Poot[1] and Hong X. Tang[1*]

**Electromagnetically induced transparency has great theoretical and experimental importance in many physics subjects, such as atomic physics, quantum optics, and more recent cavity optomechanics. Optical delay is the most prominent feature of electromagnetically induced transparency, and in cavity optomechanics optical delay is limited by mechanical dissipation rate of sideband-resolved mechanical modes. Here we demonstrate a cascaded optical transparency scheme by leveraging the parametric phonon-phonon coupling in a multimode optomechanical system, where a low damping mechanical mode in the unresolved-sideband regime is made to couple to an intermediate, high frequency mechanical mode in the resolved-sideband regime of an optical cavity. Extended optical delay and higher transmission, as well as optical advancing are demonstrated. These results provide a route to realize ultra-long optical delay, indicating a significant step toward integrated classical and quantum information storage devices.**

[1] Department of Electrical Engineering, Yale University, 15 Prospect St., New Haven, Connecticut 06511, USA
[*] e-mail: hong.tang@yale.edu

It is well known that photon is the ideal choice for long-distance communication for its low propagation loss, high speed and large bandwidth. In both quantum and classical domain, optical delay is a favored feature for advanced optical networks, as it offers the ability to buffer and store optical signals[1]. One promising method to produce optical delay is through electromagnetically induced transparency (EIT) [2], which arises from the quantum interference between different excitation paths as demonstrated in many platforms[3,4,5,6,7]. The optical transmission and delay time are two important parameters to describe the performance of one EIT process. For example, quantum memories have been proposed based on EIT effect[8,9], where the optical transmission and delay directly determine the information transfer efficiency and storage time[10,11]. EIT effect has also been demonstrated with cavity optomechanical systems[6,7], which has advanced into the quantum regime, and paved the way to realize integrated quantum devices[12,13,14,15,16]. In spite of great advances in the performance of optomechanical devices, it remains challenging in typical cavity optomechanical systems to realize EIT effect with both high optical transmission and long optical delay. The resolved-sideband condition requires high mechanical frequency which inevitably leads to high mechanical dissipation rate, thus small optical delay[6,7]. Also due to transparency window broadening under large optomechanical cooperativity[6,7], the optical transmission can only be increased with the expense of further decreased optical delay.

In this article, we propose and experimentally demonstrate the cascaded EIT effect utilizing both optomechanical coupling and parametric phonon-phonon coupling[17], in which the optical transmission and optical delay can be improved simultaneously. An aluminum nitride (AlN) wheel structure is fabricated to support whispering gallery optical modes, a low frequency breathing mechanical mode and a high frequency pinch mechanical mode. The parametric phonon-phonon coupling is first characterized by electromagnetically induced transparency and absorption effect in the subsystem consisting of the two mechanical modes. Then by combining the optomechanical coupling and parametric phonon-phonon coupling, a secondary narrow transparency window produced by the low frequency breathing mechanical mode is observed on the top of the original wide transparency window produced by the high frequency pinch mechanical mode. In this cascaded EIT scheme, the optical delay is improved by 8 times with simultaneous increase in optical transmission compared with normal EIT configuration.

**Results**

**Theoretical description.** As shown in Fig. 1a and b, we start with a regular optomechanical system in the resolved-sideband regime, where an optical cavity (frequency $\omega_o$, linewidth $\kappa$) is coupled to a high frequency mechanical mode (frequency $\Omega_H$, linewidth $\Gamma_H$) with coupling rate $g_{om}$. We introduce another sideband-unresolved low frequency mechanical mode (frequency $\Omega_L$, linewidth $\Gamma_L$) which couples to the high frequency mechanical mode through parametric phonon-phonon coupling with coupling rate $g_{mm}$. The interaction Hamiltonian of the parametric phonon-phonon coupling has a similar form as radiation pressure coupling in optomechanical systems (Supplementary Note 1). In cavity optomechanical systems, the photon-phonon coupling can be strongly enhanced by a coherent optical pump, leading to EIT phenomena. Analogously the subsystem consisting of the two parametrically coupled mechanical modes should also exhibit EIT phenomena[17].

We use the following Hamiltonian to describe the coupled multimode system:

$$H = \hbar\omega_o a^+ a + \hbar\Omega_H b^+ b + \hbar\Omega_L c^+ c + \hbar g_{om} a^+ a (b^+ + b) + \hbar g_{mm} b^+ b (c^+ + c). \qquad (1)$$

Here $a$, $b$, and $c$ are the annihilation operators of the optical cavity, high and low frequency mechanical modes respectively, and $g_{om}$ ($g_{mm}$) represents the zero-point optomechanical (parametric phonon-phonon) coupling rate. In the presence of an optical pump (which results in a coherent optical field $\alpha_0$) at the red side of the optical cavity and a mechanical pump ($\beta_0$) at the blue side of the high frequency mechanical mode, the interaction Hamiltonian is found by substituting $a \to (\alpha_0 + a)$, $b \to (\beta_0 + b)$ under the rotating wave approximation (RWA)

$$H_i = \hbar g_{om}(\alpha_0^+ ab^+ + \alpha_0 a^+ b) + \hbar g_{mm}(\beta_0 b^+ c^+ + \beta_0^+ bc). \tag{2}$$

If a weak optical probe is swept across the optical cavity, the first term in equation (2) causes destructive interference between the probe photons and the pump photons scattered by the high frequency mechanical mode phonons, inducing transparency in the transmitted probe light[6,7]. The second term causes constructive interference between the high frequency mechanical mode phonons and the mechanical pump scattered by the low frequency mechanical mode phonons. The energy level diagram of this cascaded EIT scheme is shown in Fig. 1c. There are three pathways for the optical probe in the cavity: i) directly transmitting through the optical cavity, ii) interfering with the optical pump scattered by the high frequency mechanical mode, iii) interfering with the optical pump scattered by the phonons scattered from the mechanical pump by the low frequency mechanical mode. The total transmitted signal of the optical probe can be written as (Supplementary Note 2)

$$t(\nu) = 1 - \frac{\kappa_{ex}}{i(\omega_o - \omega_p - \nu) + \frac{\kappa}{2} + \frac{g_{om}^2 |\alpha_0|^2}{i(\Omega_H - \nu) + \frac{\Gamma_H}{2} - \frac{g_{mm}^2 |\beta_0|^2}{-i(\Omega_L - \Omega_p + \nu) + \Gamma_L/2}}}, \tag{3}$$

where $\omega_p$ and $\Omega_p$ are the optical and mechanical pump frequencies respectively, $\kappa_{ex}$ is the optical coupling loss, and $\nu$ is the frequency difference between the optical pump and probe (Fig.

1c). Beside the transparency window produced due to the high frequency mechanical mode, feature of the low frequency mechanical mode is also imprinted in the optical probe through the parametric phonon-phonon coupling with the high frequency mechanical mode serving as a bridge. To observe the cascaded EIT, the resolved-sideband condition is required individually for the optomechanical coupling ($\Omega_H > \kappa$) and parametric phonon-phonon coupling ($\Omega_L > \Gamma_H$), but does not restrict the low frequency mechanical mode to be in the resolved-sideband regime with respect to the optical cavity. Therefore, it can be designed to have lower frequency which offers a much smaller linewidth than the high frequency mechanical mode, and leads to prolonged group delay (Fig. 1d).

**Device design and characterization.** The parametric phonon-phonon coupling is a key part in the cascaded EIT, which we implement in a micro-wheel structure (Fig. 2a). A displacement of one mechanical mode leads to tension and boundary change in the structure, thus modifying the resonant frequencies of other mechanical modes. Unlike the coupling induced by the Duffing effect in flexural beams where the frequency shift is proportional to the square of the displacement[18,19], the frequency shift for the radial modes in micro-wheel structures is linearly proportional to the displacement because of the broken mirror symmetry in the radial direction. The coupling rate in equations (1) and (2) can be expressed as (Supplementary Note 1):

$$g_{mm} = \frac{\int_V (\epsilon_H^T:\mathbf{C}:\epsilon_H)(\nabla \cdot \mathbf{u}_L)dV + 2\int_V (\epsilon_H^T:\mathbf{C}:\epsilon_L)(\nabla \cdot \mathbf{u}_H)dV}{2\Omega_H \int_V \rho |\mathbf{u}_H|^2 dV} \cdot u_{L,\text{zpm}}. \qquad (4)$$

Here $\mathbf{u}_H$ and $\mathbf{u}_L$ are the normalized mode profiles, $\epsilon_H = \frac{1}{2}[\nabla \mathbf{u}_H + (\nabla \mathbf{u}_H)^T]$ and $\epsilon_L = \frac{1}{2}[\nabla \mathbf{u}_L + (\nabla \mathbf{u}_L)^T]$ are the strain tensors, $\mathbf{C}$ is the fourth-order elasticity tensor, $\rho$ is the material density, and $u_{L,\text{zpm}}$ is the zero-point motion of the low frequency mode. The first term accounts for the potential energy change of the high frequency mode due to the low frequency mode displacement, and the second term accounts for the fact that the two mechanical modes are not orthogonal when structure deformation is considered. Note that the parametric phonon-phonon coupling is an intrinsic property in multimode mechanical devices, unlike the coupling induced by external electric and optical fields[20,21,22].

The efficient piezoelectric drive of mechanical resonators makes multimode cavity piezo-optomechanical systems an ideal platform to realize our cascaded EIT[23,24]. We use an on-chip AlN wheel structure to obtain high-$Q$ optical and mechanical modes, as well as high piezoelectric driving efficiency[23,24] (Fig. 2b). As an added benefit, the piezoelectric effect in AlN further enhances the parametric phonon-phonon coupling by effectively stiffening the material (Supplementary Note 1). The device optimization and fabrication process are shown in Supplementary Note 3. The optical whispering gallery mode around 1561.94 nm wavelength has a loaded $Q$ of 380,000, corresponding to $\kappa = 2\pi \times 0.51$ GHz (Fig. 2c). The mechanical modes are characterized by piezoelectrically actuating the device and optically detecting the response in a liquid helium cryostat at temperature $T = 4.6$ K (Fig. 2d, e). The two mechanical modes used in the experiment are the breathing mode ($\Omega_L = 2\pi \times 76.29$ MHz, $\Gamma_L = 2\pi \times 2.7$ kHz) and the pinch mode ($\Omega_H = 2\pi \times 1.094$ GHz, $\Gamma_H = 2\pi \times 33.6$ kHz), and the corresponding zero-point phonon-phonon coupling rate is $2\pi \times 0.028$ Hz. This coupling rate can be greatly enhanced by a coherent mechanical pump which could provide a large number of phonons. The pinch mode has a frequency significantly larger than the linewidth of the optical cavity, and the same holds for the breathing mode frequency compared to the linewidth of the pinch mode, so both satisfy their respective resolved-sideband requirement. Also the breathing mode frequency is significantly small compared with the optical cavity linewidth, thus in the optomechanical unresolved-sideband regime. The optomechanical coupling between the optical mode and the low frequency breathing mode is negligible under the RWA due to the frequency mismatch.

**Parametric phonon-phonon coupling.** We first examine the EIT in the subsystem consisting of the two mechanical modes induced by the parametric phonon-phonon coupling. No optical pump is used, and a weak mechanical probe ($\beta_1$) with frequency $\nu$ is swept across the high frequency mechanical mode, whose motion is detected using a weak optical readout. Without phonon-phonon coupling, one simply measures the Lorentzian driven response of the high frequency mode as shown in Fig. 2 (e). With the co-existing strong mechanical pump ($\beta_0$) and weak mechanical probe ($\beta_1$), the radiation-like force due to the parametric phonon-phonon coupling introduces phonons into the low frequency mechanical mode, as it drives the resonator at the frequency difference between $\beta_0$ and $\beta_1$. These phonons in turn scatter the mechanical pump, which interfere with the mechanical probe. This EIT effect results in the sharp dip at $\nu = \Omega_H$ in Fig. 3a when the mechanical pump is placed at $\Omega_p = \Omega_H - \Omega_L$. The five panels in Fig. 3b show the normalized mechanical probe signal when the mechanical pump frequency is varied from $\Omega_H - \Omega_L - 2\Gamma_H$ to $\Omega_H - \Omega_L + 2\Gamma_H$. The transparency window always occurs at frequency $\nu = \Omega_p + \Omega_L$ with a maximum transparency window achieved at $\Omega_p = \Omega_H - \Omega_L$. When the mechanical pump is placed at the blue side of the high frequency mechanical mode, an absorption peak manifests itself indicating an increased phonon number in the high frequency mechanical mode (Fig. 3c).

**Cascaded optical transparency.** By combining the parametric phonon-phonon coupling and optomechanical coupling, the cascaded EIT effect can be observed. When the optical probe is swept across the optical cavity over a wide range, we first see the normal EIT which has a transparency window with a linewidth equal to the dissipation rate of the high frequency mechanical mode (Fig. 4a). Upon zooming into the transparency window, a second transparency window sitting on top of the original transparency window can be observed (Fig. 4b). In this case, the linewidth of the narrow transparency window equals to the dissipation rate of the low frequency mechanical mode instead of the much larger linewidth of the high frequency mode. The high frequency mechanical mode essentially mediates the coupling between the optical cavity and the low frequency mechanical mode which is in the optomechanical unresolved-sideband regime and therefore cannot be used for regular EIT. By fitting the spectrum in Fig. 4b with equation (3), we can obtain both the optomechanical cooperativity $C_{\text{om}} = \frac{4g_{\text{om}}^2|\alpha_0|^2}{\kappa \Gamma_H} \approx 0.15$, and the phonon-phonon coupling cooperativity $C_{\text{mm}} = \frac{4g_{\text{mm}}^2|\beta_0|^2}{\Gamma_H \Gamma_L} \approx 0.29$. Both cooperativities are close to unity allowing the observation of the cascaded EIT. The narrow optical transparency window leads to strong optical phase modulations as shown in Fig. 4 (b) and therefore large optical group delays, which in our case reach 5.0 μs. This corresponds to a more than 8 times improvement compared with 0.6 μs delay of the regular EIT which only involves the high frequency mode (Fig. 4d). Moreover, the optical transmission is increased, and this leads to a 10 times improvement in the bandwidth-delay product[9]. By placing the mechanical pump on the red side of the high frequency mechanical mode, we turn the narrow peak into a dip (Fig. 4c). This gives rise to the advanced light with a leading time of 1.03 μs (Fig. 4d).

**Discussion**

For optomechanical applications in information processing, one major obstacle is the short phonon lifetime which introduces noise during state transfer and limits the information storage time after state transfer[25,26,27,28]. In our optomechanical system, the phonon lifetime for the high frequency mechanical mode is around 30 $\mu s$. The phonon lifetime can be greatly increased in the cascaded multimode coupling scheme. In this scheme, the low frequency mechanical mode in the optomechanical unresolved-sideband regime functions as if it is in the optomechanical resolved-sideband regime, and the frequency just needs to exceed the linewidth of the high frequency mechanical mode (~33 kHz). For a mechanical resonator with frequency around 200 kHz, a linewidth as low as 0.1 Hz can be realized, corresponding to a 10 s phonon lifetime[29]. This increase in phonon lifetime is directly proportional to the information storage time increase for a classical signal. Also a higher mechanical $Q$ can be expected for a lower frequency mechanical mode[30] due to the empirical $f \cdot Q$ limit[31]. Thereforethe low frequency mechanical mode can preserve the quantum state much longer due to its longer rethermalization time $\tau = \hbar Q/kT$[32], thus serving as a better quantum register[33]. In essence, our cascaded coupling scheme provides a viable route to long optical delay and long coherence time by leveraging the parametric phonon-phonon coupling, pushing forward the frontier of optomechanical devices in both classical and quantum regime.

## Methods

**Measurement Setup.** The experiment is realized with the setup shown in Fig. 2g. In the experiment demonstrating parametric phonon-phonon coupling (Fig. 3), the RF signal from a network analyzer (NA) goes through port 2 of the RF switch (SW), and is sent into the device under test (DUT) to serve as the weak mechanical probe ($\beta_1$). The mechanical pump ($\beta_0$) is provided by the amplified RF signal from a signal generator (SG). No optical pump is used in this part of the experiment, and only a fixed weak readout laser is delivered to the DUT to linearly transduce the mechanical signal. The transmitted optical signal is amplified by an erbium-doped fiber amplifier (EDFA), and detected by a photoreceiver (PR). Then the electrical signal is sent back to the NA.

In the experiment demonstrating cascaded EIT (Fig. 4), the optical pump ($\alpha_0$) from a laser is delivered to the DUT after passing through an electro-optic modulator (EOM). The RF signal from the NA goes through port 1 of the switch to modulate the strong optical pump, and this generates a weak optical probe ($\alpha_{p+}$). In this case, the only RF signal sent into the device is the mechanical pump ($\beta_0$) which is provided by the amplified RF signal from SG. The transmitted optical signal is amplified by the EDFA, and detected by the photoreceiver. Then the signal is sent back to the NA.

### Author contributions

L. F. performed the experiment and data analysis with the support of K. Y. F. and supervision by H. X. T. The interpretation and modelling were done by L. F., K. Y. F., and M. P. All authors contribute to the writing of the manuscript.

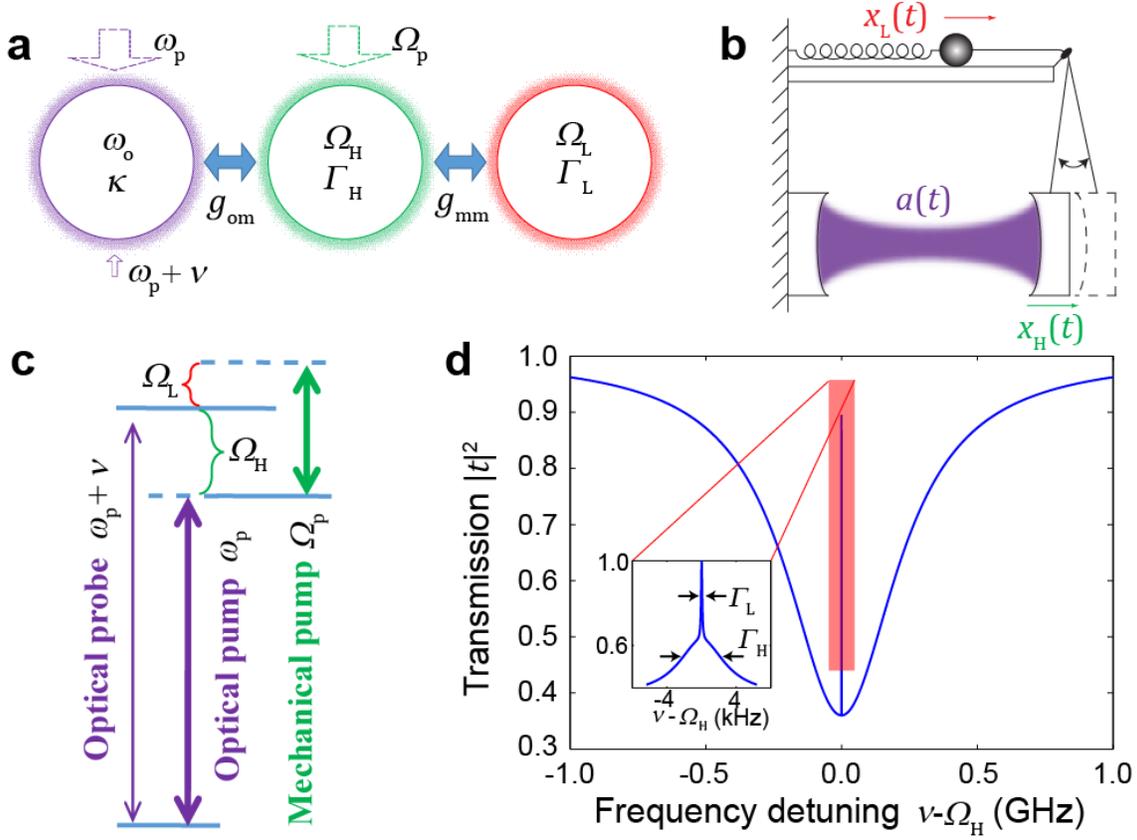

**Figure 1 | Cascaded EIT Model. a,** Schematic of the multimode coupling scheme. The optical and mechanical pumps are positioned at $\omega_p$ and $\Omega_p$ respectively, and a weak optical probe is located at $\omega_p + \nu$. **b,** A model of the multimode coupling. One mirror of the Fabry-Perot cavity (the optical cavity) is suspended from a pivot, forming a pendulum (the high frequency mechanical mode), and the length of the pendulum (hence the frequency) is determined by the position of a small ball attached to a spring (the low frequency mechanical mode). **c,** The energy-level diagram of the multimode coupling scheme. The optical pump frequency is $\omega_p = \omega_o - \Omega_H$, and the mechanical pump frequency is $\Omega_p = \Omega_H + \Omega_L$. **d,** Calculated transmission spectrum of the probe light (equation (3)) as a function of the probe light frequency $\nu$. The inset reveals the ultra-narrow cascaded transparency window induced by the multimode coupling. The parameters used are $\omega_o = 2\pi \times 194$ THz, $\kappa = 2\pi \times 0.5$ GHz, $\kappa_{ex} = 0.2\kappa$; $\Omega_H = 2\pi \times 1$ GHz, $\Gamma_H = 2\pi \times 20$ kHz; $\Omega_L = 2\pi \times 80$ MHz, $\Gamma_L = 2\pi \times 2$ kHz; $g_{om}|\alpha_0| = 2\pi \times 1.5$ MHz, $g_{mm}|\beta_0| = 2\pi \times 3$ kHz.

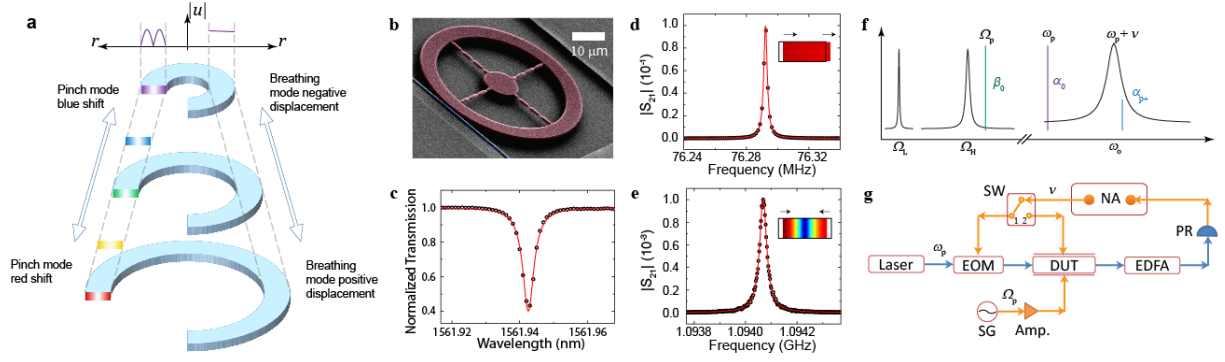

**Figure 2 | Experimental system for studying cascaded EIT. a,** Parametric phonon-phonon coupling mechanism with pinch and breathing modes in a wheel structure. The breathing mode is depicted in the deformed shapes with three different displacements. The pinch mode is depicted as the cross-section profile (the saturation corresponds to the radial displacement amplitude). The decrease and increase of the radius cause opposite frequency changes of the pinch mode. **b,** SEM image of the device in false color. The inner and outer radius of the AlN wheel are 20 μm and 25 μm respectively. **c,** Transmission spectrum of the optical whispering gallery mode at 1561.94 nm. **d,** Driven response of the breathing mode. **e,** Driven reponse of the pinch mode. The insets in **d** and **e** show the displacement profiles of the wheel cross-section for each mechanical mode. **f,** Configuration of the coherent signals, optical and mechanical resonances in the experiment. **g,** The measurement setup. NA: Network Analyzer, SW: RF switch, SG: Signal Generator, Amp: Amplifier, DUT: Device Under Test, EDFA: Erbium-Doped Fiber Amplifier, PR: Photo-Receiver, EOM: Electro-Optic Modulator.

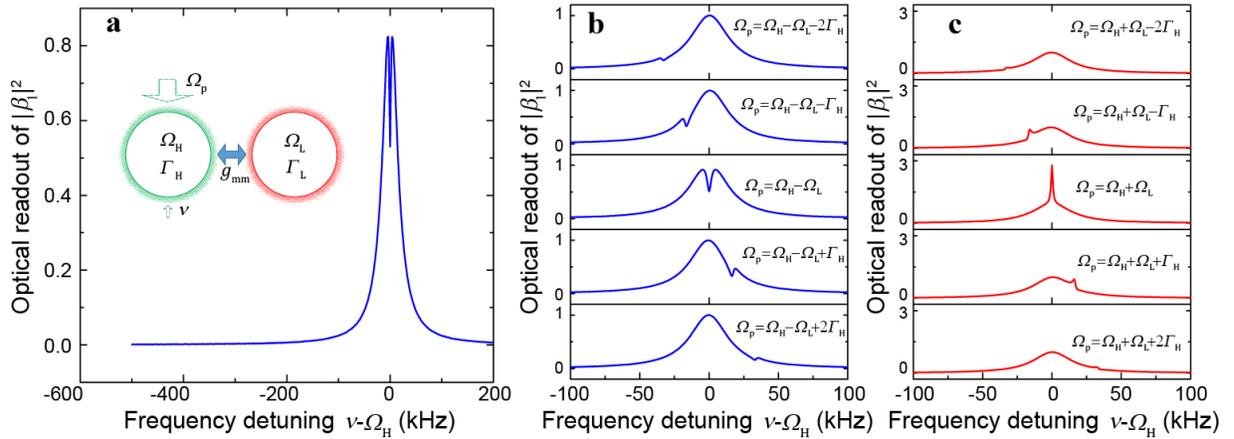

**Figure 3 | Parametric phonon-phonon coupling in the subsystem consisting of the two mechanical modes. a,** The optically transduced high frequency mechanical probe ($|\beta_1|^2$) when the mechanical pump is placed at $\Omega_p = \Omega_H - \Omega_L$. The sharp dip at $v = \Omega_H$ characterizes the EIT phenomenon induced by parametric phonon-phonon coupling between the two mechanical modes. **b,** Zoom-in of the transparency window with adjusted detuning on the red side of the high frequency mechanical mode. **c,** Zoom-in of the absorption window with mechanical pump detuned on the blue side of the high frequency mechanical mode.

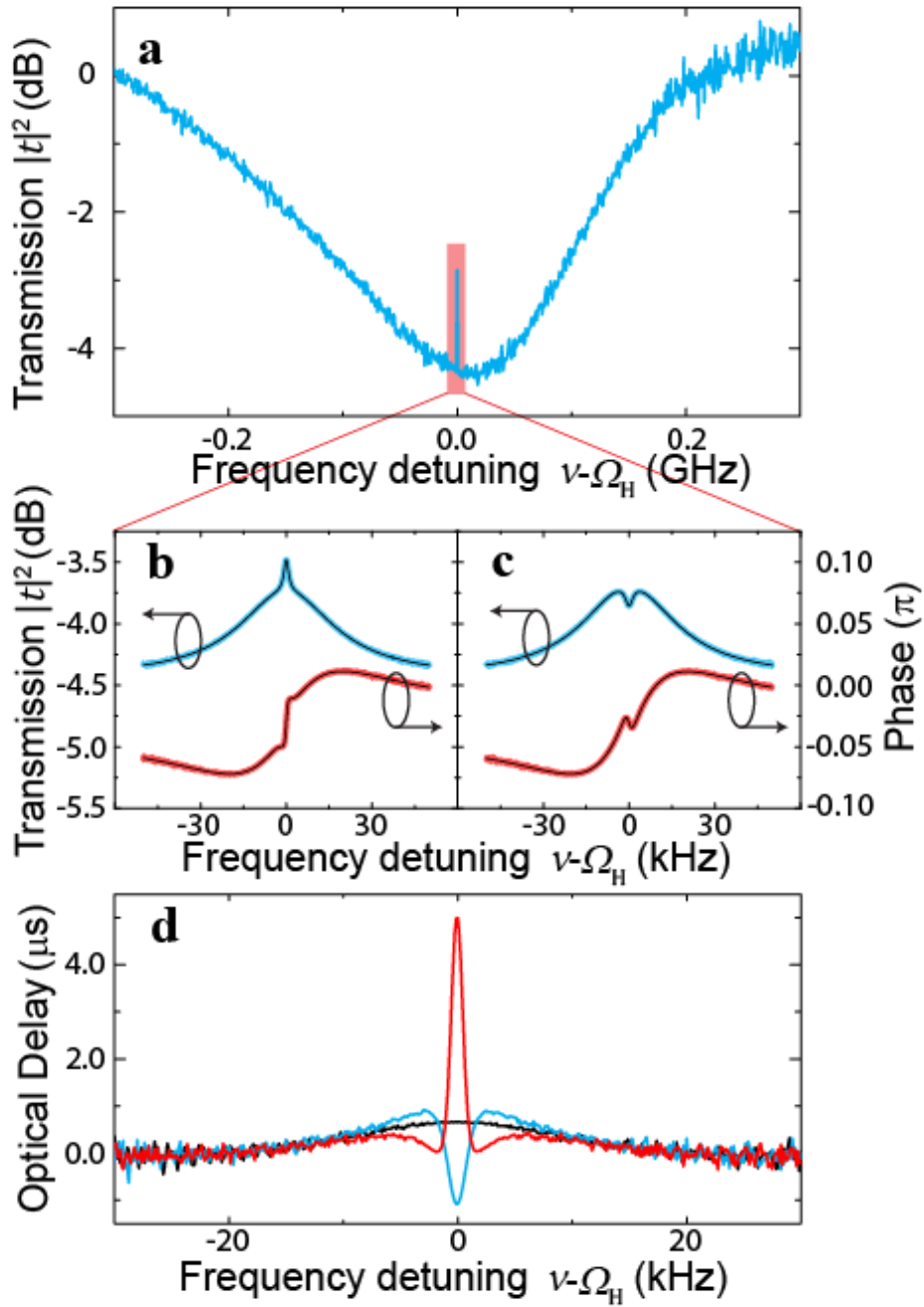

**Figure 4 | Cascaded EIT a,** The probe light transmission ($|\alpha_{p+}|^2$) over the broad frequency range. **b,** Zoom-in of the transmission (cyan) and phase (red) of the transparency window when the mechanical pump is placed on the blue side ($\Omega_p = \Omega_H + \Omega_L$). This is the configuration for the prolonged optical delay. **c,** The zooming-in transmission (cyan) and phase (red) of the transparency window when the mechanical pump is placed on the red side ($\Omega_p = \Omega_H - \Omega_L$). This

is the configuration for the enhanced optical advancing. The black lines in **b** and **c** are the best fitting curves of our cascaded EIT theory. **d,** The optical delay for the normal EIT configuration (black), cascaded EIT with the prolonged optical delay configuration as in **b** (red), and cascaded EIT with the enhanced optical advancing configuration as in **c** (cyan)

# Supplementary Information

## Supplementary Figures

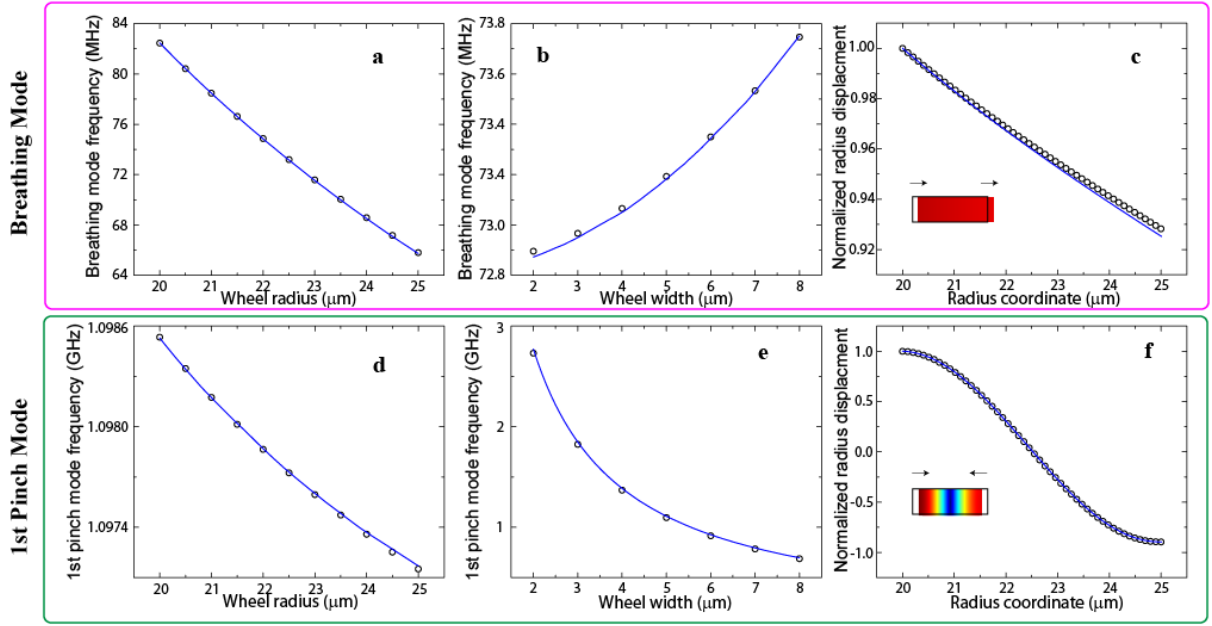

**Supplementary Figure 1 Mechanical mode calculation**. The resonant frequency of the breathing mode is plotted versus the wheel radius $\frac{R_1+R_2}{2}$ in **a**, and wheel width $R_1 - R_2$ in **b**. The mode profile along the radius direction is also presented in **c** with $R_1 = 20$ μ2 and $R_2 = 25$ μ2. The resonant frequency of the 1$^{st}$ pinch mode is plotted versus the wheel radius $\frac{R_1+R_2}{2}$ in **d**, and wheel width $R_1 - R_2$ in **e**. The mode profile along the radius direction is also presented in **f** with $R_1 = 20$ μ2 and $R_2 = 25$ μ2. In all figures, the blue line is calculated based on the theory derived (Eq. (5) – (8)), and the black circles are FEM simulation results. The material parameters used are $\rho = 3300$ kg/m$^3$, $C_{11} = 410$ GPa, $C_{12} = 149$ GPa, $C_{13} = 99$ GPa, $C_{33} = 389$ GPa, $C_{44} = 125$ GPa[1] , $d_{15} = -3.84$ pm/V, $d_{31} = -1.91$ pm/V, $d_{33} = 4.96$ pm/V[2]. Here $C_{ij}$ are the elements of the elasticity matrix **C**, which is the inverse of compliance matrix.

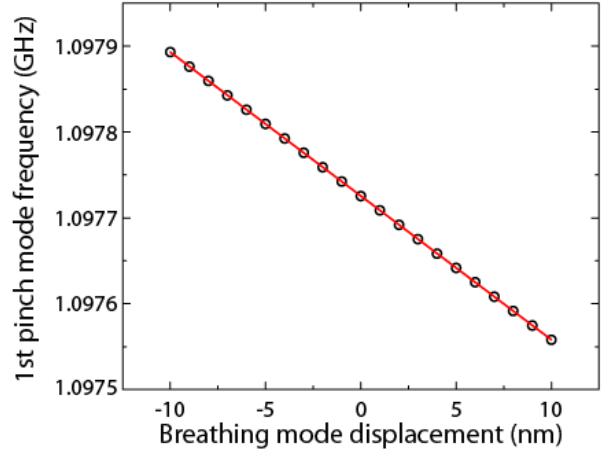

**Supplementary Figure 2 Parametric phonon-phonon coupling rate calculation.** The FEM simulated dependence between the 1$^{st}$ pinch mode frequency and breathing mode displacement. The linear fit gives the phonon-phonon coupling rate of $G_{mm} = -2\pi \times 0.017$ MHz/nm. The parameters used in simulation is the same as Supplementary Figure 1.

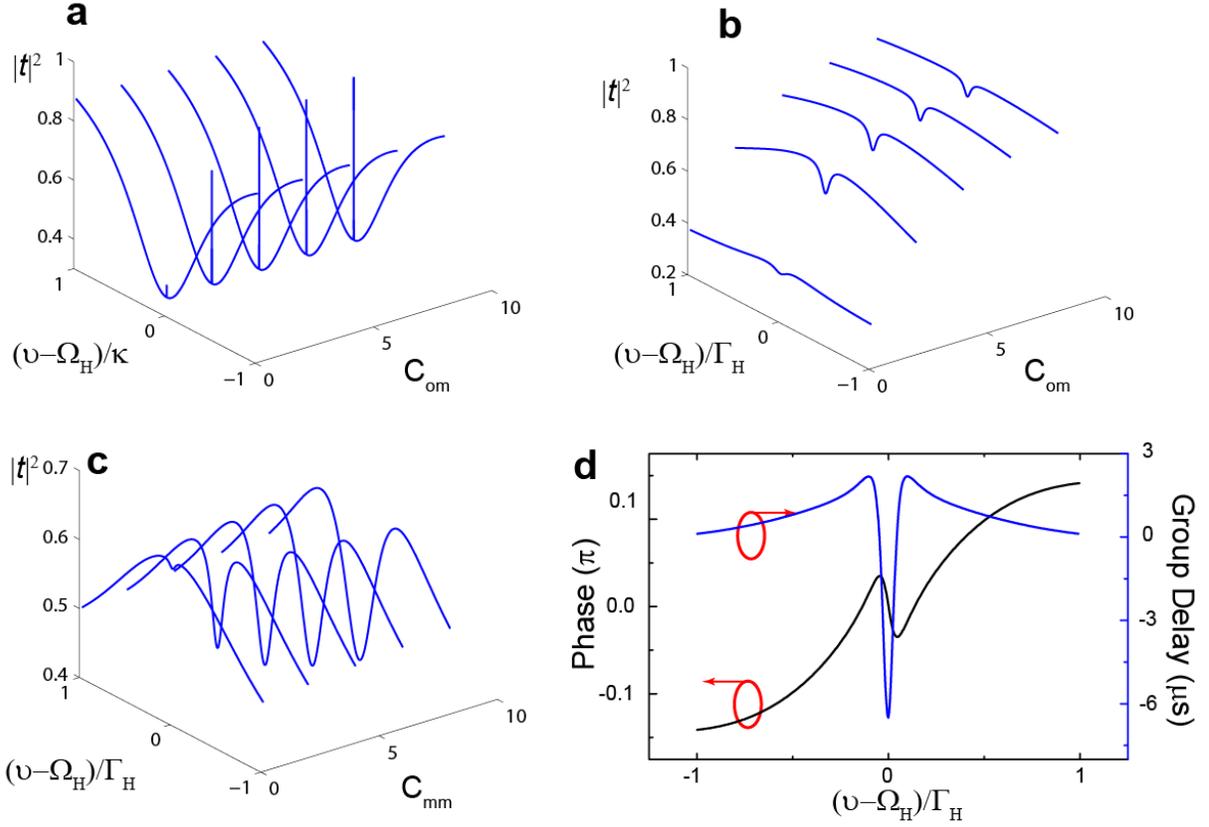

**Supplementary Figure 3** The calculated optical probe transmission $|t|^2$ under red-detuned optical pump and red-detuned mechanical pump as a function of $\nu$ with different optomechanical cooperativity $C_{om}$ and fixed phonon-phonon cooperativity $C_{mm} = 1$ over large and small frequency range in **a** and **b** respectively. **c,** $|t|^2$ in Eq. (28) as a function of $\nu$ with fixed optomechanical cooperativity $C_{om} = 1$ and different phonon-phonon cooperativity $C_{mm}$. **d,** The phase change and group delay with $C_{om} = 1$ and $C_{mm} = 1$. The parameters used are $\omega_o = 2\pi \times 194$ THz, $\kappa = 2\pi \times 0.5$ GHz, $\kappa_{ex} = 0.2\kappa$; $\Omega_H = 2\pi \times 1$ GHz, $\Gamma_H = 2\pi \times 30$ kHz; $\Omega_L = 2\pi \times 76$ MHz, $\Gamma_L = 3$ kHz; $\omega_p = \omega_o - \Omega_H, \Omega_p = \Omega_H - \Omega_L$.

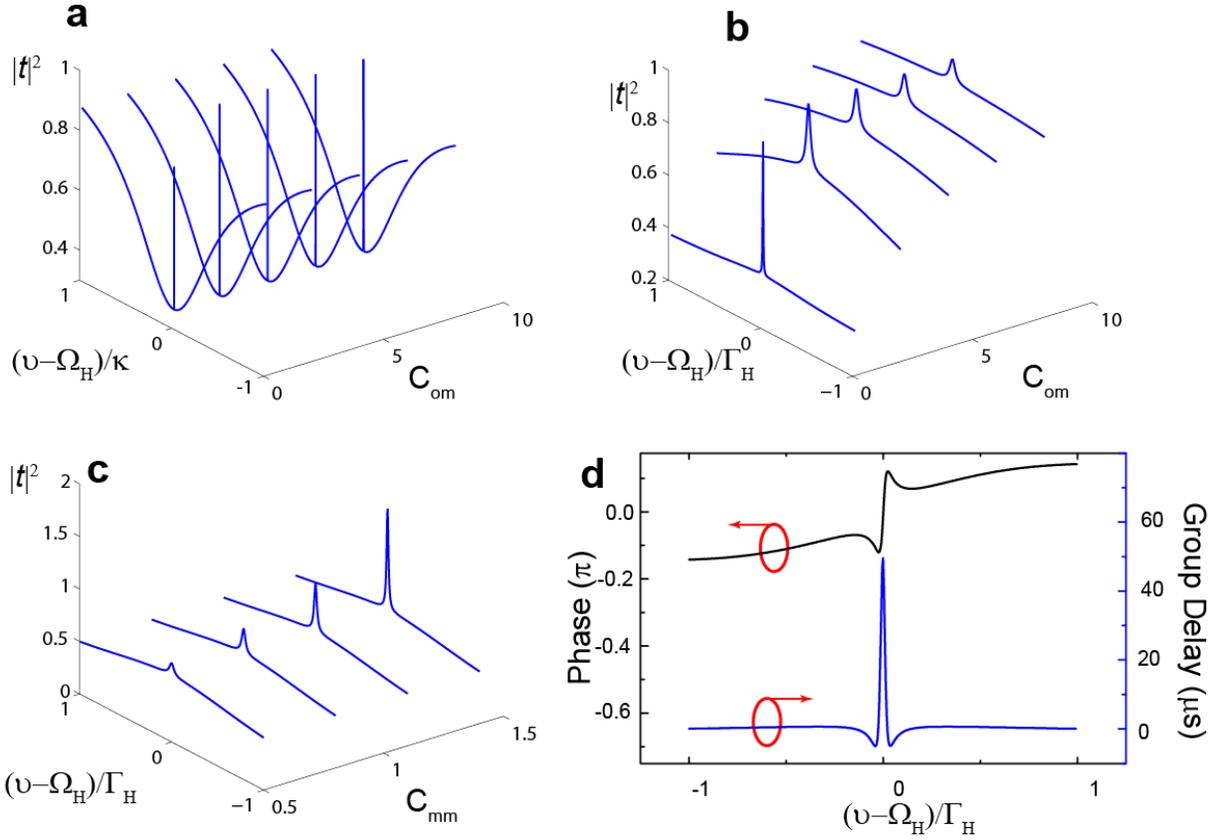

**Supplementary Figure 4** The calculated optical probe transmission $|t|^2$ under red-detuned optical pump and blue-detuned mechanical pump as a function of $\nu$ with different optomechanical cooperativity $C_{om}$ and fixed phonon-phonon cooperativity $C_{mm} = 1$ over large and small frequency range in **a** and **b** respectively. **c,** $|t|^2$ in Eq. (30) as a function of $\nu$ with fixed optomechanical cooperativity $C_{om} = 1$ and different phonon-phonon cooperativity $C_{mm}$. **d,** The phase change and group delay with $C_{om} = 1$ and $C_{mm} = 1$. The parameters used are $\omega_o = 2\pi \times 194$ THz, $\kappa = 2\pi \times 0.5$ GHz, $\kappa_{ex} = 0.2\kappa$; $\Omega_H = 2\pi \times 1$ GHz, $\Gamma_H = 2\pi \times 30$ kHz; $\Omega_L = 2\pi \times 76$ MHz, $\Gamma_L = 3$ kHz; $\omega_p = \omega_o - \Omega_H$, $\Omega_p = \Omega_H + \Omega_L$.

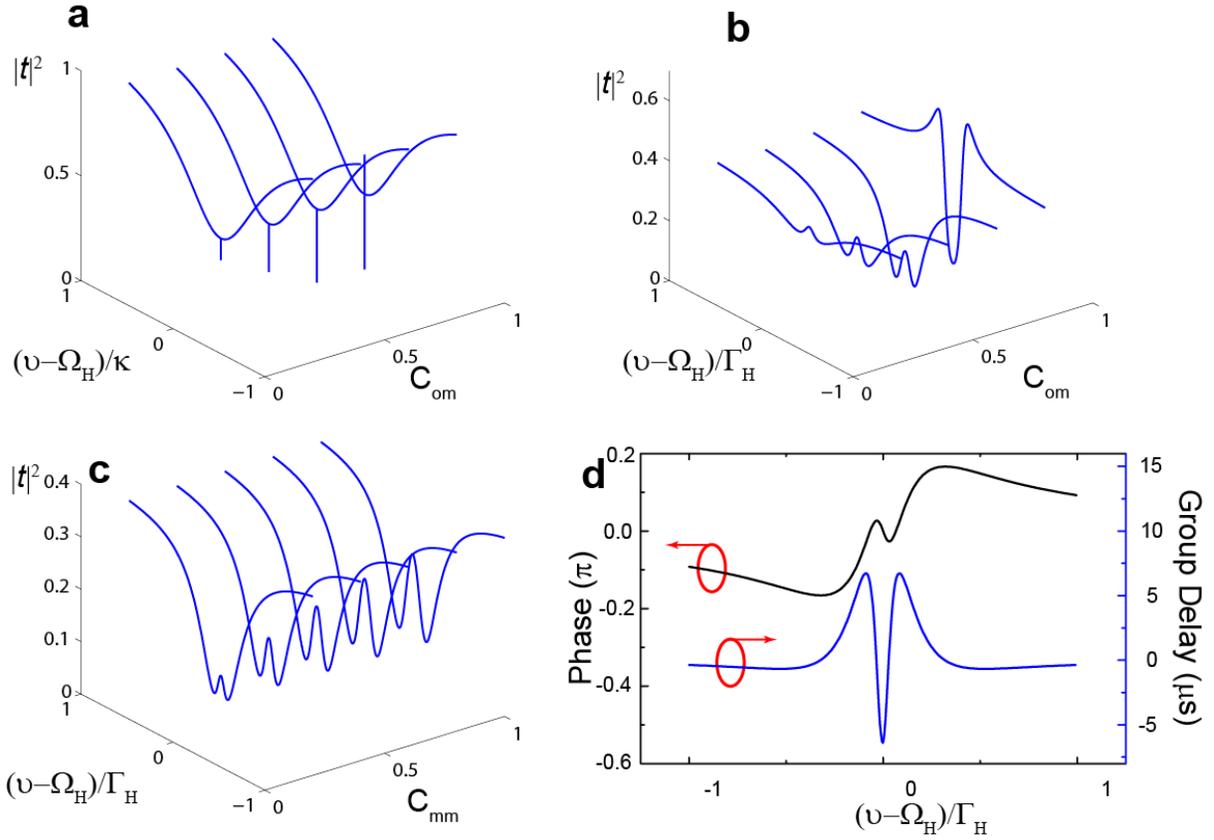

**Supplementary Figure 5** The calculated optical probe transmission $|t|^2$ under blue-detuned optical pump and red-detuned mechanical pump as a function of $\nu$ with different optomechanical cooperativity $C_{om}$ and fixed phonon-phonon cooperativity $C_{mm} = 0.5$ over large and small frequency range in **a** and **b** respectively. **c,** $|t|^2$ in Eq. (31) as a function of $\nu$ with fixed optomechanical cooperativity $C_{om} = 0.5$ and different phonon-phonon cooperativity $C_{mm}$. **d,** The phase change and group delay with $C_{om} = 0.3$ and $C_{mm} = 0.3$. The parameters used are $\omega_o = 2\pi \times 194$ THz, $\kappa = 2\pi \times 0.5$ GHz, $\kappa_{ex} = 0.2\kappa$; $\Omega_H = 2\pi \times 1$ GHz, $\Gamma_H = 2\pi \times 30$ kHz; $\Omega_L = 2\pi \times 76$ MHz, $\Gamma_L = 3$ kHz; $\omega_p = \omega_o + \Omega_H, \Omega_p = \Omega_H - \Omega_L$.

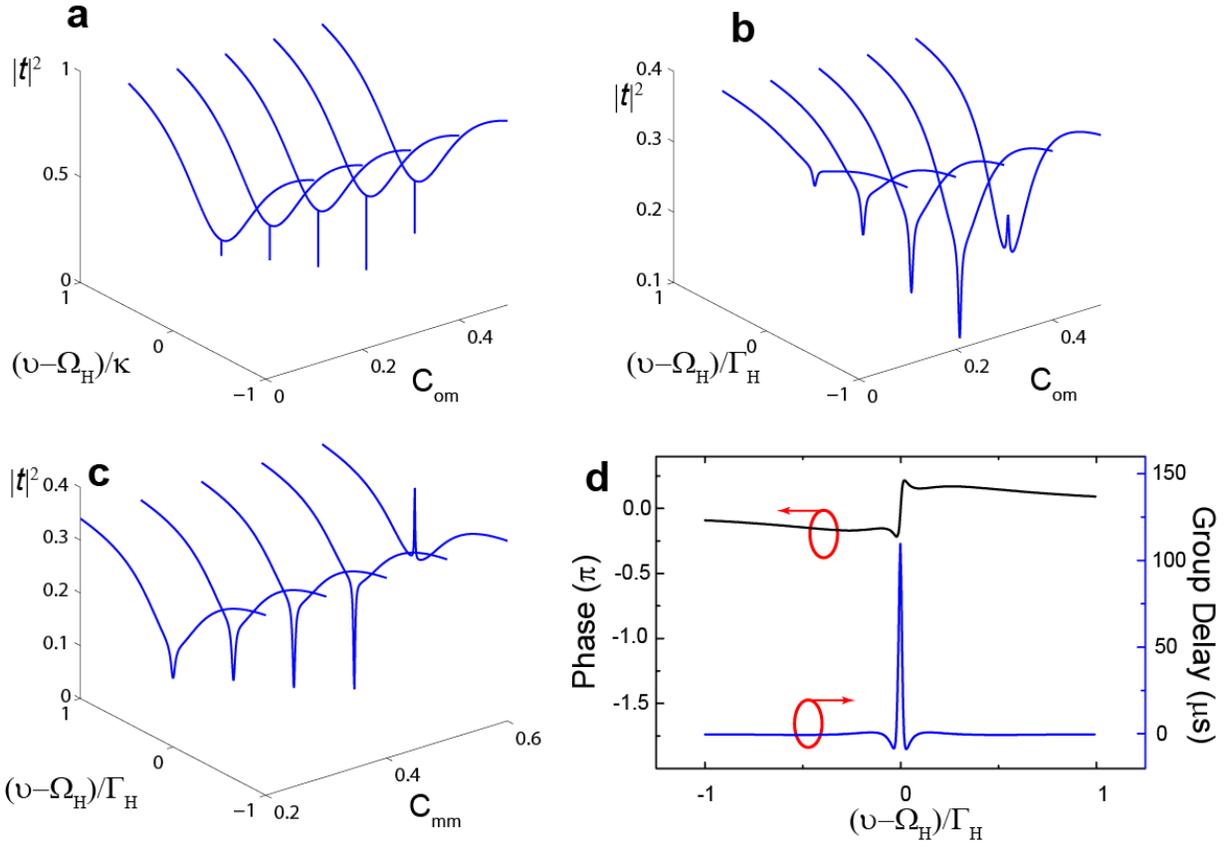

**Supplementary Figure 6** The calculated optical probe transmission $|t|^2$ under blue-detuned optical pump and blue-detuned mechanical pump as a function of $\nu$ with different optomechanical cooperativity $C_{om}$ and fixed phonon-phonon cooperativity $C_{mm} = 0.3$ in large and small frequency range in **a** and **b** respectively. **c,** $|t|^2$ in Eq. (32) as a function of $\nu$ with fixed optomechanical cooperativity $C_{om} = 0.3$ and different phonon-phonon cooperativity $C_{mm}$. **d,** The phase change and group delay with $C_{om} = 0.3$ and $C_{mm} = 0.3$. The parameters used are $\omega_o = 2\pi \times 194$ THz, $\kappa = 2\pi \times 0.5$ GHz, $\kappa_{ex} = 0.2\kappa$; $\Omega_H = 2\pi \times 1$ GHz, $\Gamma_H = 2\pi \times 30$ kHz; $\Omega_L = 2\pi \times 76$ MHz, $\Gamma_L = 3$ kHz; $\omega_p = \omega_o + \Omega_H, \Omega_p = \Omega_H + \Omega_L$.

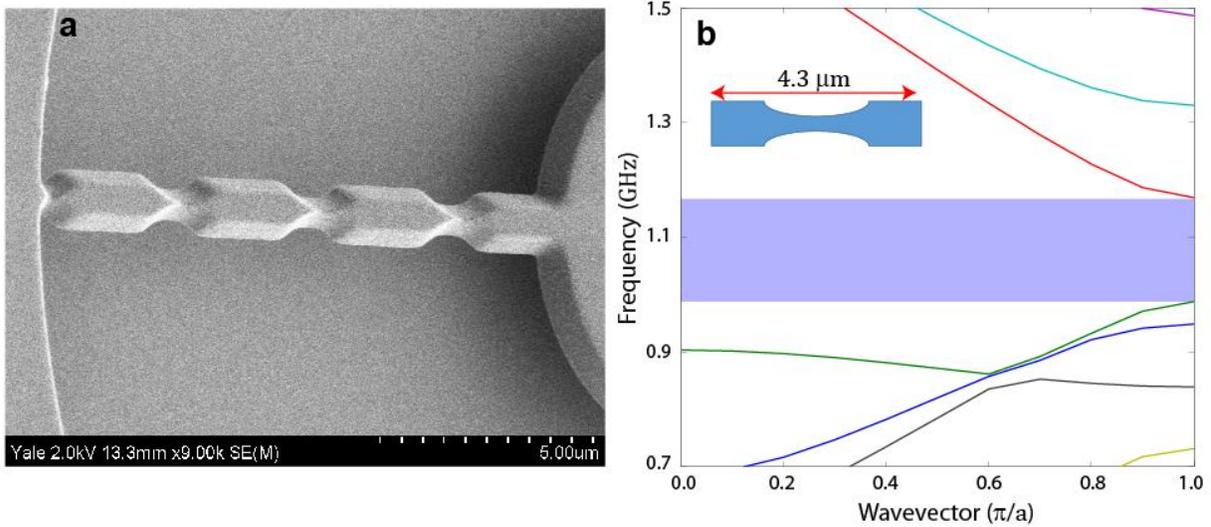

**Supplementary Figure 7 Phononic crystal engineering. a.** SEM picture of phononic crystal structure on spokes. **b.** The band structure of phononic crystal on spokes, the shaded area corresponds to the full bandgap between 0.99 GHz and 1.17 GHz. The inset shows the top view of the phononic crystal unit cell.

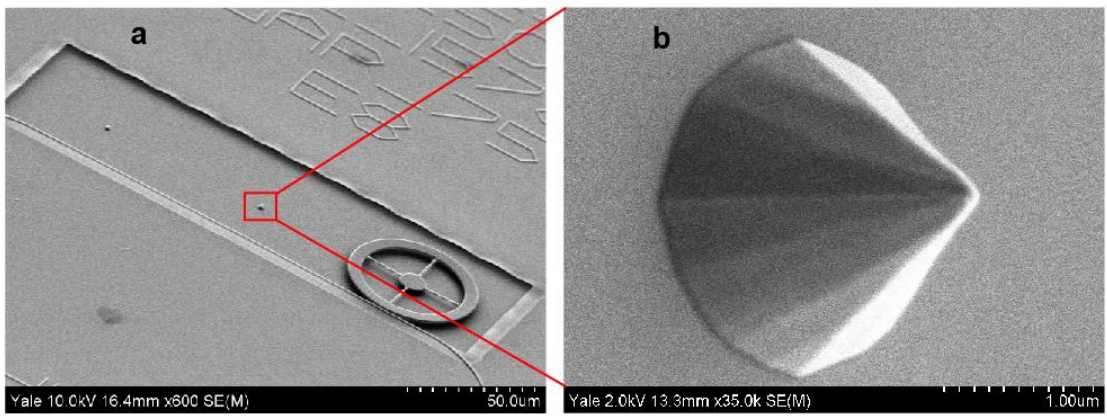

**Supplementary Figure 8 Device undercut engineering a.** SEM picture of a device where the two smaller devices in the same releasing window disappeared after BOE wet etch. **b.** SEM picture of a pedestal which belongs to a disappeared device. The pedestal which belongs to the existing device should have a similar size.

# Supplementary Note 1. Theory of dispersive phonon-phonon coupling

# Mechanical mode resonant frequency and profile

Here we provide an analytical method to calculate profiles and resonant frequencies of mechanical modes with rotation symmetry in AlN wheels. The crystal structure of AlN is wurtzite structure, therefore the constitutive equations can be expressed as

$$\begin{pmatrix} \epsilon_{rr} \\ \epsilon_{\theta\theta} \\ \epsilon_{zz} \\ \epsilon_{z\theta} \\ \epsilon_{zr} \\ \epsilon_{r\theta} \\ D_r \\ D_\theta \\ D_z \end{pmatrix} = \begin{pmatrix} s_{11}^E & s_{12}^E & s_{13}^E & 0 & 0 & 0 & 0 & 0 & d_{31} \\ s_{12}^E & s_{11}^E & s_{13}^E & 0 & 0 & 0 & 0 & 0 & d_{31} \\ s_{13}^E & s_{13}^E & s_{33}^E & 0 & 0 & 0 & 0 & 0 & d_{33} \\ 0 & 0 & 0 & s_{44}^E & 0 & 0 & 0 & d_{15} & 0 \\ 0 & 0 & 0 & 0 & s_{44}^E & 0 & d_{15} & 0 & 0 \\ 0 & 0 & 0 & 0 & 0 & s_{66}^E & 0 & 0 & 0 \\ 0 & 0 & 0 & 0 & d_{15} & 0 & \varepsilon_{11} & 0 & 0 \\ 0 & 0 & 0 & d_{15} & 0 & 0 & 0 & \varepsilon_{11} & 0 \\ d_{31} & d_{31} & d_{33} & 0 & 0 & 0 & 0 & 0 & \varepsilon_{33} \end{pmatrix} \times \begin{pmatrix} \sigma_{rr} \\ \sigma_{\theta\theta} \\ \sigma_{zz} \\ \sigma_{z\theta} \\ \sigma_{zr} \\ \sigma_{r\theta} \\ E_r \\ E_\theta \\ E_z \end{pmatrix}, \tag{1}$$

where $\epsilon_{rr}, \epsilon_{\theta\theta}, \epsilon_{zz}, \epsilon_{z\theta}, \epsilon_{zr}, \epsilon_{r\theta}$ are the components of the strain tensor, $D_r, D_\theta, D_z$ are the components of the electric displacement, $\sigma_{rr}, \sigma_{\theta\theta}, \sigma_{zz}, \sigma_{z\theta}, \sigma_{zr}, \sigma_{r\theta}$ are the components of the stress tensor, $E_r, E_\theta, E_z$ are the components of the electric field. $s_{ij}^E$ denote the compliance components at constant electric field, $d_{ij}$ denote the piezoelectric constants, and $\varepsilon_{ij}$ are the dielectric constants at constant stress.

All the electric displacement components equal zero, because there is no free charge on the device. Also we only consider the radial displacement $u(r,t)$, because mechanical modes have rotation symmetry and the wheel thickness is very small[3]. Therefore, the constitutive equations can be simplified to

$$\epsilon_{rr} = \frac{\partial u(r,t)}{\partial r}, \tag{2a}$$

$$\epsilon_{\theta\theta} = \frac{u(r,t)}{r}, \tag{2b}$$

$$\sigma_{rr} = \frac{1}{s_{11}^E(1-v_p^2)}(\epsilon_{rr} + v_p \epsilon_{\theta\theta}) - \frac{d_{31}}{s_{11}^E(1-v_p)} E_z, \tag{2c}$$

$$\sigma_{\theta\theta} = \frac{1}{s_{11}^E(1-v_p^2)}(v_p \epsilon_{rr} + \epsilon_{\theta\theta}) - \frac{d_{31}}{s_{11}^E(1-v_p)} E_z, \tag{2d}$$

$$D_z = d_{31}(\sigma_{rr} + \sigma_{\theta\theta}) + \varepsilon_{33} E_z = 0, \tag{2e}$$

where $v_p = -\frac{s_{21}^E}{s_{11}^E}$ is the planar Poisson's ratio. The only remaining equation needed is the force equation

$$\rho \frac{\partial^2 u(r,t)}{\partial t^2} = \frac{\partial \sigma_{rr}}{\partial r} + \frac{\sigma_{rr} - \sigma_{\theta\theta}}{r}, \tag{3}$$

where $\rho$ is the material density. Substituting Eq. (2) into Eq. (3) and assuming harmonic motion with angular frequency $\Omega$ (i.e. $u(r,t) = u(r)e^{-i\Omega t}$), we obtain the equation of motion expressed by the radial displacement

$$\frac{\partial^2 u}{\partial r^2} + \frac{1}{r} \cdot \frac{\partial u}{\partial r} - \frac{u}{r^2} + (1-v_p^2)\rho \Omega^2 s_{11}^{EP} u = 0, \tag{4}$$

where $s_{11}^{EP} = \frac{s_{11}^{E}}{1+\frac{k_p^2(1+v_p)}{2(1-k_p^2)}}$, and $k_p = \sqrt{\frac{2d_{31}^2}{s_{11}^{E}(1-v_p)\varepsilon_{33}}}$ is the planar electromechanical coupling coefficient[4]. The piezoelectric effect changes the material compliance from $s_{11}^{E}$ to $s_{11}^{EP}$, and $k_p$ indicates the strength of the piezoelectric effect. The solution of Eq. (4) is

$$u = AJ_1(p\Omega r) + BY_1(p\Omega r), \tag{5}$$

where $p = \sqrt{\rho(1-v_p^2)s_{11}^{EP}}$, and $J_1$ and $Y_1$ are the first-order Bessel functions of the first and second kind, and $A$ and $B$ are the constants determined by the boundary conditions. The boundary conditions are $\sigma_{rr} = 0$ at $r = R_1$ and $r = R_2$ with $R_1$ and $R_2$ representing the inner and outer radius of the wheel.

$$AG_1(p\Omega R_1) + BG_2(p\Omega R_1) = 0, \tag{6a}$$
$$AG_1(p\Omega R_2) + BG_2(p\Omega R_2) = 0, \tag{6b}$$

where $G_1$ and $G_2$ are defined as the following

$$G_1(x) = x\left(1 + \frac{\alpha(1+v_p)}{2(1-\alpha)}\right)J_0(x) + (v_p - 1)J_1(x), \tag{7a}$$

$$G_2(x) = x\left(1 + \frac{\alpha(1+v_p)}{2(1-\alpha)}\right)Y_0(x) + (v_p - 1)Y_1(x), \tag{7b}$$

where $J_0$ and $Y_0$ are the zero-order Bessel functions of the first and second kind. In order to get non-trivial values for $A$ and $B$, the equation's determinant must be zero

$$\begin{vmatrix} G_1(p\Omega R_1) & G_2(p\Omega R_1) \\ G_1(p\Omega R_2) & G_2(p\Omega R_2) \end{vmatrix} = 0. \tag{8}$$

By solving Eq. (8), we can get a series of solutions corresponding to different resonant frequencies (Supplementary Figure 1). And the ratio between $A$ and $B$ can be obtained from Eq. (6a), thus the mode profile can be plotted from Eq. (5) (Supplementary Figure 1).

## Dispersive phonon-phonon coupling

In this section, we use Lagrangian mechanics method to derive the coupling between two mechanical modes when considering the structure deformation induced by the mechanical mode displacement. We first derive the equation for arbitrary structures. We consider two mechanical modes with displacement $\mathbf{x}_1 = \mathbf{u}_1(\mathbf{r})A_1(t)$ and $\mathbf{x}_2 = \mathbf{u}_2(\mathbf{r})A_2(t)$, where $\mathbf{u}_1$ and $\mathbf{u}_2$ are the normalized mode profiles ($\max(|\mathbf{u}_{1,2}|) = 1$). The corresponding strain fields are

$$\boldsymbol{\epsilon}_1(\mathbf{r})A_1(t) = \frac{A_1(t)}{2}[\nabla \mathbf{u}_1 + (\nabla \mathbf{u}_1)^T], \tag{9a}$$

$$\boldsymbol{\epsilon}_2(\mathbf{r})A_2(t) = \frac{A_2(t)}{2}[\nabla \mathbf{u}_2 + (\nabla \mathbf{u}_2)^T], \tag{9b}$$

where $(\cdot)^T$ represents a transpose. The total system kinetic energy ($T$) is

$$T = \frac{1}{2}\int_V \rho \left|\frac{\partial(\mathbf{x}_1+\mathbf{x}_2)}{\partial t}\right|^2 dV = \frac{\dot{A}_1^2}{2}\int_V \rho|\mathbf{u}_1|^2 dV + \frac{\dot{A}_2^2}{2}\int_V \rho|\mathbf{u}_2|^2 dV. \tag{10}$$

And the potential energy ($PE$) can be written as

$$PE = \frac{1}{2}\int_V [(\boldsymbol{\epsilon}_1 A_1 + \boldsymbol{\epsilon}_2 A_2)^T : \mathbf{C} : (\boldsymbol{\epsilon}_1 A_1 + \boldsymbol{\epsilon}_2 A_2)](1 + \nabla \cdot \mathbf{u}_1 A_1 + \nabla \cdot \mathbf{u}_2 A_2) dV, \quad (11)$$

where $\mathbf{C}$ is the fourth-order elasticity tensor, and $\mathbf{f} : \mathbf{g} \equiv \sum_{ij} f_{ij} g_{ij}$ is the second-order inner product. Here the term $(1 + \nabla \cdot \mathbf{u}_1 A_1 + \nabla \cdot \mathbf{u}_2 A_2)$ accounts for the structure deformation. Then we follow the standard procedure for Lagrangian mechanics

$$\frac{\partial (T-PE)}{\partial A_1} - \frac{d}{dt}\left(\frac{\partial (T-PE)}{\partial \dot{A}_1}\right) = 0. \quad (12)$$

We plug Eq. (10), Eq. (11) and Eq. (12) into Eq. (13), and only keep the 1st and 2nd order of $A_1$ and $A_2$. We further ignore the self-coupling terms ($A_1^2$, $A_2^2$). Then we get

$$\left[\int_V \rho |\mathbf{u}_1|^2 dV\right] \ddot{A}_1 + \left[\begin{array}{c} \int_V (\boldsymbol{\epsilon}_1^T : \mathbf{C} : \boldsymbol{\epsilon}_1) dV + A_2 \int_V (\boldsymbol{\epsilon}_1^T : \mathbf{C} : \boldsymbol{\epsilon}_1)(\nabla \cdot \mathbf{u}_2) dV \\ + 2 A_2 \int_V (\boldsymbol{\epsilon}_1^T : \mathbf{C} : \boldsymbol{\epsilon}_2)(\nabla \cdot \mathbf{u}_1) dV \end{array}\right] A_1 = 0. \quad (13)$$

If we define the following parameters:

$$M_1 = \int_V \rho |\mathbf{u}_1|^2 dV, \quad (14a)$$

$$\omega_{1,0}^2 = \frac{\int_V (\boldsymbol{\epsilon}_1^T : \mathbf{C} : \boldsymbol{\epsilon}_1) dV}{M_1}, \quad (14b)$$

$$K_1 = \frac{\int_V (\boldsymbol{\epsilon}_1^T : \mathbf{C} : \boldsymbol{\epsilon}_1)(\nabla \cdot \mathbf{u}_2) dV}{M_1}, \quad (14c)$$

$$K_2 = \frac{2 \int_V (\boldsymbol{\epsilon}_1^T : \mathbf{C} : \boldsymbol{\epsilon}_2)(\nabla \cdot \mathbf{u}_1) dV}{M_1}, \quad (14d)$$

then Eq. (13) can be simplified to

$$\ddot{A}_1 + \left(\omega_{1,0}^2 + (K_1 + K_2) A_2\right) A_1 = 0. \quad (15)$$

Here $M_1$ and $\omega_{1,0}$ are the effective mass and resonant frequency of the 1st mechanical mode respectively, $K_1$ and $K_2$ are the coupling coefficients caused by the structure deformation. The effect of $K_1$ and $K_2$ is to change the frequency to $\omega_1^2 = \omega_{1,0}^2 + (K_1 + K_2) A_2$. The phonon-phonon coupling rate thus can be written as

$$G_{\mathrm{mm}} = \frac{d\omega_1}{dA_2} = \frac{K_1 + K_2}{2\omega_1} = \frac{\int_V (\boldsymbol{\epsilon}_1^T : \mathbf{C} : \boldsymbol{\epsilon}_1)(\nabla \cdot \mathbf{u}_2) dV + 2 \int_V (\boldsymbol{\epsilon}_1^T : \mathbf{C} : \boldsymbol{\epsilon}_2)(\nabla \cdot \mathbf{u}_1) dV}{2\omega_1 \int_V \rho |\mathbf{u}_1|^2 dV} \quad \mathrm{Hz/m}. \quad (16)$$

And the interaction Hamiltonian is expressed as

$$H_{\mathrm{mi}} = \hbar G_{\mathrm{mm}} u_{2,\mathrm{zpm}} b^+ b (c^+ + c). \quad (17)$$

where $b$ ($c$) is the annihilation operator of the 1st (2nd) mechanical mode, and $u_{2,\mathrm{zpm}}$ is the zero-point motion of the 2nd mechanical mode. When the structure deformation is small, Eq. (15) is simplified to normal harmonic oscillator equation

$$\ddot{A}_1 + \omega_{1,0}^2 A_1 = 0. \quad (18)$$

Next we calculate the phonon-phonon coupling rate of two mechanical modes in the AlN wheel. We use Voigt notation in the following derivation for simplicity. Following the assumptions in the last section, we only consider mechanical modes with rotation symmetry, thus the total displacement can be written as

$$(u, v, w) = (u_1(r)A_1(t) + u_2(r)A_2(t), 0, 0) \tag{19}$$

where $u_1(r)$ and $u_2(r)$ are the normalized displacement profiles along radial direction, and $A_1(t)$ and $A_2(t)$ are the amplitudes of each mode. The strain field is

$$\begin{pmatrix} \epsilon_{rr} \\ \epsilon_{\theta\theta} \\ \epsilon_{zz} \\ \epsilon_{z\theta} \\ \epsilon_{zr} \\ \epsilon_{r\theta} \end{pmatrix} = \begin{pmatrix} \frac{\partial u}{\partial r} \\ \frac{u}{r} \\ 0 \\ 0 \\ 0 \\ 0 \end{pmatrix} = \begin{pmatrix} \frac{du_1}{dr} A_1(t) + \frac{du_2}{dr} A_2(t) \\ \frac{u_1}{r} A_1(t) + \frac{u_2}{r} A_2(t) \\ 0 \\ 0 \\ 0 \\ 0 \end{pmatrix}. \tag{20}$$

Therefore the kinetic energy ($T$) of the system is

$$T = \frac{\rho}{2} \int_0^{2\pi} \int_{R_1}^{R_2} \left( u_1 \frac{dA_1}{dt} + u_2 \frac{dA_2}{dt} \right)^2 h \cdot r d\theta dr = \pi \rho h \int_{R_1}^{R_2} (u_1 \dot{A}_1 + u_2 \dot{A}_2)^2 r dr. \tag{21}$$

The potential energy ($PE$), including both the elastic energy and electrical energy, can be written as

$$PE = \frac{1}{2} \int_V (\boldsymbol{\epsilon}^T : \mathbf{C} : \boldsymbol{\epsilon} + \mathbf{E}^T \cdot \varepsilon \cdot \mathbf{E}) dV$$

$$= \frac{1}{2} \iiint (C_{11}^P \epsilon_{rr}^2 + C_{11}^P \epsilon_{\theta\theta}^2 + 2C_{12}^P \epsilon_{rr} \epsilon_{\theta\theta}) \left( 1 + \left(\frac{u_1}{r} + \frac{du_1}{dr}\right) A_1 + \left(\frac{u_2}{r} + \frac{du_2}{dr}\right) A_2 \right) r dr dz d\theta$$

$$= \pi h \int_{R_1}^{R_2} \left[ C_{11}^P \left( \frac{du_1}{dr} A_1 + \frac{du_2}{dr} A_2 \right)^2 + C_{11}^P \left( \frac{u_1}{r} A_1 + \frac{u_2}{r} A_2 \right)^2 + 2C_{12}^P \left( \frac{du_1}{dr} A_1 + \frac{du_2}{dr} A_2 \right) \left( \frac{u_1}{r} A_1 + \frac{u_2}{r} A_2 \right) \right] \cdot$$
$$\left[ r + \left( u_1 + r \frac{du_1}{dr} \right) A_1 + \left( u_2 + r \frac{du_2}{dr} \right) A_2 \right] dr. \tag{22}$$

where $\mathbf{E}$ is the electric field in the structure, $h$ is the wheel thickness, and $C_{11}^P = C_{11} + \frac{\varepsilon_{33} k_p^4}{4 d_{31}^2 (1 - k_p^2)^2}$, $C_{21}^P = C_{12} + \frac{\varepsilon_{33} k_p^4}{4 d_{31}^2 (1 - k_p^2)^2}$. Thus the parameters in Eq. (14) in this case are expressed as

$$M_1 = 2\pi \rho h \int_{R_1}^{R_2} u_1^2 r dr, \tag{23a}$$

$$\omega_{1,0}^2 = \frac{2\pi h \int_{R_1}^{R_2} \left[ C_{11}^P \left( \frac{du_1}{dr} \right)^2 + C_{11}^P \left( \frac{u_1}{r} \right)^2 + 2C_{12}^P \left( \frac{u_1 du_1}{r \, dr} \right) \right] r dr}{M_1}, \tag{23b}$$

$$K_1 = \frac{2\pi h \int_{R_1}^{R_2} \left[ C_{11}^P \left( \frac{du_1}{dr} \right)^2 + C_{11}^P \left( \frac{u_1}{r} \right)^2 + 2C_{12}^P \left( \frac{u_1 du_1}{r \, dr} \right) \right] \left( u_2 + r \frac{du_2}{dr} \right) dr}{M_1}, \tag{23c}$$

$$K_2 = \frac{4\pi h \int_{R_1}^{R_2} \left[ C_{11}^P \left( \frac{du_1 du_2}{dr \, dr} \right) + C_{11}^P \left( \frac{u_1 u_2}{r^2} \right) + C_{12}^P \left( \frac{1 d(u_1 u_2)}{r \, dr} \right) \right] \left( u_1 + r \frac{du_1}{dr} \right) dr}{M_1}. \tag{23d}$$

Considering the wheel used in the experiment, we have the geometric parameters $R_1 = 20$ μm, $R_2 = 25$ μm, $h = 650$ nm, and the material parameters are the same as those in Supplementary Figure 1. If we use the 1st pinch mode as $u_1$ and the breahing mode as $u_2$ whose mode profiles are given in the last section, then we can get $K = K_1 + K_2 = -2\pi \times 2.64 \times 10^{23}$ Hz$^2$/m, corresponding to the coupling rate $G_{mm} = -2\pi \times 0.019$ MHz/nm. With FEM simulation, the dependence between the 1$^{st}$ pinch mode frequency and breathing mode displacement can be obtained (Supplementary Figure 2), and the simulated phonon-phonon coupling rate is $-2\pi \times 0.017$ MHz/nm which agrees well with the theoretical calculation.

## Supplementary Note 2. Theory of cascaded transparency

We propose to use cascaded EIT to realize ultra-narrow optical transparency window, utilizing both optomechanical coupling and parametric phonon-phonon coupling. The system is modeled with the Hamiltonian

$$H = \hbar\omega_o a^+ a + \hbar\Omega_H b^+ b + \hbar\Omega_L c^+ c + \hbar g_{om} a^+ a(b^+ + b) + \hbar g_{mm} b^+ b(c^+ + c) \quad (24)$$

With the optical pump ($a_{in}$), optical probe ($a_{in+}$ and $a_{in-}$, generated by modulating the optical pump with electro-optic modulator), and mechanical pump ($b_{in}$), the equations of motion can be written as

$$\dot{a} = -\left(i\omega_o + \frac{\kappa}{2}\right)a - ig_{om}a(b + b^+) + \sqrt{\kappa_{ex}}\left(a_{in}e^{-i\omega_p t} + a_{in+}e^{-i(\omega_p+\nu)t} + a_{in-}e^{-i(\omega_p-\nu)t}\right), \quad (25a)$$

$$\dot{b} = -\left(i\Omega_H + \frac{\Gamma_H}{2}\right)b - ig_{om}a^+ a - ig_{mm}b(c + c^+) + \sqrt{\Gamma_{H,ex}}b_{in}e^{-i\Omega_p t}, \quad (25b)$$

$$\dot{c} = -\left(i\Omega_L + \frac{\Gamma_L}{2}\right)c - ig_{mm}b^+ b. \quad (25c)$$

We solve Eq. (25) in the steady state, and make the following substitution.

$$a = \alpha_0 e^{-i\omega_p t} + \alpha_{p+} e^{-i(\omega_p+\nu)t} + \alpha_{p-} e^{-i(\omega_p-\nu)t} + \alpha_+ e^{-i(\omega_p+\Omega_p)t} + \alpha_- e^{-i(\omega_p-\Omega_p)t}, \quad (26a)$$

$$b = \beta_s + \beta_0 e^{-i\Omega_p t} + \beta_1 e^{-i\nu t}, \quad (26b)$$

$$c = \gamma_s + \gamma_1 e^{-i|\nu-\Omega_p|t}. \quad (26c)$$

We assume the optical pump and mechanical pump are very strong so that the amplitudes of $\alpha_0$ and $\beta_0$ are determined by the optical pump and mechanical pump respectively, and the effect of static displacement $\beta_s$ and $\gamma_s$ can be eliminated by a frequency renormalization. Thus the equations for optical sidebands and mechanical motions are

$$\left[i(\omega_o - \omega_p - \nu) + \frac{\kappa}{2}\right]\alpha_{p+} = \sqrt{\kappa_{ex}}a_{in+} - ig_{om}\alpha_0\beta_1, \quad (27a)$$

$$\left[i(\omega_o - \omega_p + \nu) + \frac{\kappa}{2}\right]\alpha_{p-} = \sqrt{\kappa_{ex}}a_{in-} - ig_{om}\alpha_0\beta_1^*, \quad (27b)$$

$$\left[i(\Omega_H - \nu) + \frac{\Gamma_H}{2}\right]\beta_1 = -ig_{om}(\alpha_0 \alpha_{p-}^* + \alpha_0^* \alpha_{p+}) - ig_{mm}\beta_0\gamma_1^* \quad (\Omega_p > \nu), \quad (27c)$$

$$\left[i(\Omega_H - \nu) + \frac{\Gamma_H}{2}\right]\beta_1 = -ig_{om}(\alpha_0 \alpha_{p-}^* + \alpha_0^* \alpha_{p+}) - ig_{mm}\beta_0\gamma_1 \quad (\Omega_p < \nu), \quad (27d)$$

$$\left[i(\Omega_L - \Omega_p + \nu) + \frac{\Gamma_L}{2}\right]\gamma_1 = -ig_{mm}\beta_0\beta_1^* \quad (\Omega_p > \nu), \quad (27e)$$

$$\left[i(\Omega_L + \Omega_p - \nu) + \frac{\Gamma_L}{2}\right]\gamma_1 = -ig_{mm}\beta_0^*\beta_1 \quad (\Omega_p < \nu). \quad (27f)$$

Here we ignore the case $\Omega_p = \nu$, in which the spectrum is a Dirac delta function because of the mixing between the mechanical and optical pumps[5]. Based on the red or blue detuning of the optical pump and mechanical pump with respect to the optical cavity and the high frequency mechanical mode, there are four different configurations in the system.

**Case I**: The optical pump is red detuned and the mechanical pump is red detuned. In this case, only $\alpha_{p+}$ has a significant contribution to the probe transmission, thus the optical probe transmission is

$$t(\nu) = 1 - \frac{\kappa_{\text{ex}}}{i(\omega_o-\omega_p-\nu)+\frac{\kappa}{2}+\dfrac{g_{\text{om}}^2|\alpha_0|^2}{i(\Omega_H-\nu)+\dfrac{\Gamma_H}{2}+\dfrac{g_{\text{mm}}^2|\beta_0|^2}{i(\Omega_L+\Omega_p-\nu)+\Gamma_L/2}}}. \tag{28}$$

This equation is plotted in Supplementary Figure 3. The transmitted light can exhibit group advancing (i.e. negative delay) which is only possible in the reflected light for normal EIT with optomechanical systems[6]. Also we can increase the transmission by only increasing the optomechanical cooperativity ($C_{\text{om}} = \frac{4g_{\text{om}}^2|\alpha_0|^2}{\kappa \Gamma_H}$), and keep the phonon-phonon cooperativity ($C_{\text{mm}} = \frac{4g_{\text{mm}}^2|\beta_0|^2}{\Gamma_L \Gamma_H}$) the same. Then, the dip's linewidth does not change significantly, and the optical advancing time does not decrease. The optical delay $\tau$ is calculated based on the phase information of the transmitted light[6]

$$\tau = Re\left\{-\frac{i}{t(\nu)}\cdot\frac{dt(\nu)}{d\nu}\right\}. \tag{29}$$

**Case II**: optical pump is red-detuned, and the mechanical pump is blue-detuned. In this case, only $\alpha_{p+}$ has a significant contribution to the probe transmission, thus the optical probe transmission is

$$t(\nu) = 1 - \frac{\kappa_{\text{ex}}}{i(\omega_o-\omega_p-\nu)+\frac{\kappa}{2}+\dfrac{g_{\text{om}}^2|\alpha_0|^2}{i(\Omega_H-\nu)+\dfrac{\Gamma_H}{2}-\dfrac{g_{\text{mm}}^2|\beta_0|^2}{-i(\Omega_L-\Omega_p+\nu)+\Gamma_L/2}}}. \tag{30}$$

This equation is plotted in Supplementary Figure 4. An ultra-narrow optical transparency window can be realized with this configuration. Note that with large mechanical cooperativity, the transmission can be larger than unity. This is due to the parametric amplification process with the blue detuned mechanical pump[7]. When the mechanical cooperativity is large enough $C_{\text{mm}} > 1 + C_{\text{om}}$, the low frequency mechanical resonator starts to self-oscillate, and the system enters the unstable regime. **Case III**: optical pump is blue-detuned, and the mechanical pump is red-detuned. In this case, only $\alpha_{p-}$ has a significant contribution to the probe transmission, thus the optical probe transmission is

$$t(\nu) = 1 - \frac{\kappa_{\text{ex}}}{i(\omega_o-\omega_p+\nu)+\frac{\kappa}{2}-\dfrac{g_{\text{om}}^2|\alpha_0|^2}{-i(\Omega_H-\nu)+\dfrac{\Gamma_H}{2}+\dfrac{g_{\text{mm}}^2|\beta_0|^2}{-i(\Omega_L+\Omega_p-\nu)+\Gamma_L/2}}}. \tag{31}$$

Eq. (31) is plotted in Supplementary Figure 5. In this configuration, we observe a narrow transparency window in the broad absorption window in the weak optomechanical coupling regime. When the optomechanical cooperativity increases, the broad absorption window turns into a transparency window, and the narrow transparency window turns into an absorption window, in which case the parametric amplification phenomena manifests[7]. **Case IV**: optical pump is blue-detuned, and the mechanical pump is blue-detuned. In this case, only $\alpha_{p-}$ has a significant contribution to the probe transmission, thus the optical probe transmission is

$$t(\nu) = 1 - \frac{\kappa_{\text{ex}}}{i(\omega_o-\omega_p+\nu)+\frac{\kappa}{2}-\dfrac{g_{\text{om}}^2|\alpha_0|^2}{-i(\Omega_H-\nu)+\dfrac{\Gamma_H}{2}-\dfrac{g_{\text{mm}}^2|\beta_0|^2}{i(\Omega_L-\Omega_p+\nu)+\Gamma_L/2}}}. \tag{32}$$

Eq. (32) is plotted in Supplementary Figure 6. In this configuration, we observe a narrow absorption window in the broad absorption window in the weak optomechanical coupling regime. The low frequency mechanical mode starts to self-oscillate when the sum of the optomechanical and phonon-phonon cooperativities exceeds unity ($C_{\text{mm}} + C_{\text{om}} > 1$), and the system enters unstable regime. One transparency peak can appear in the optical absorption window when the sum of the optomechanical and phonon-phonon cooperativities is large enough (c.f. Supplementary Figure 5), and the dramatic phase modulation can also lead to prolonged optical delay. Compare with **Case II**, we find both of these two cases can be used for prolonged optical delay, but **Case IV** has a much higher requirement for the optomechanical and phonon-phonon cooperativities to realize transparency window.

Supplementary Note 3. Device performance optimization

## Phonon bandgap engineering for mechanical pinch mode

The clamping loss of the mechanical mode is further reduced by fabricating phononic crystal structures on the spokes (Supplementary Figure 7a). The frequency of the high frequency pinch mode is around 1.094 GHz, so a period of $a = 4.3$ μm is chosen, and the unit cell consists of one block (4.3 μm long and 1 μm wide) with two half ellipse holes (400 nm short axis and 800 nm long axis) on each side as shown in the inset of Supplementary Figure 7b. This structure results in a full phononic bandgap between 0.99 GHz and 1.17 GHz (Supplementary Figure 7b). Because of the fabrication process, the sidewall of the structure is not completely vertical. Our simulation results show little influence of the tilted sidewall on the phononic bandgap structure. This phononic bandgap engineering turns out to be quite effective despite that we only employ three repeating segments in the spokes.

## Device Undercut engineering for minimizing clamping loss

The material mechanical loss can be greatly decreased by cooling the device to low temperature. As a result the total mechanical loss is determined by the clamping loss. Engineering the undercut of the device to minimize the pedestal is crucial to minimize the mechanical clamping loss[8]. In order to achieve a minimized pedestal, we pattern three devices with slightly different center disk radii (100 nm step) in the same releasing window (Supplementary Figure 8a). During the BOE wet etch step, we monitor these three devices. When the two devices with smaller center disk radiuses disappear, we stop the BOE wet etch. With this method, we can achieve pedestal sizes around 100 nm (Supplementary Figure 8b).

## Resist reflow technique for minimizing optical scattering loss

In order to improve the optical performance of the device, a resist reflow technique is utilized. The device is fabricated from AlN-on-insulator wafers with 650 nm AlN on a 3-μm layer of buried oxide. A layer of $SiO_2$ with thickness 150 nm is first deposited on the wafer through plasma-enhanced chemical vapor deposition. Then patterns are defined in ma-N 2403 resist by electron-beam lithography. After resist development, the device is placed on the hot-plate to partially melt the ma-N 2403 patterns and form a much smoother boundary in the patterns. Next, the patterns are transferred to the $SiO_2$ hardmask layer by $CHF_3$-based plasma dry etching, and then to AlN layer by $Cl_2$-based plasma dry etching in which only 400 nm out of 650 nm AlN is initially etched. Then a releasing window is defined with a second electron-beam lithography step using ZEP520A resist, followed by dry etching of the residual AlN layer. Last, the underlying oxide in the exposed windows is removed by buffered oxide etchant (BOE). The mechanical structures are dried in a critical point dryer.

## Supplementary References